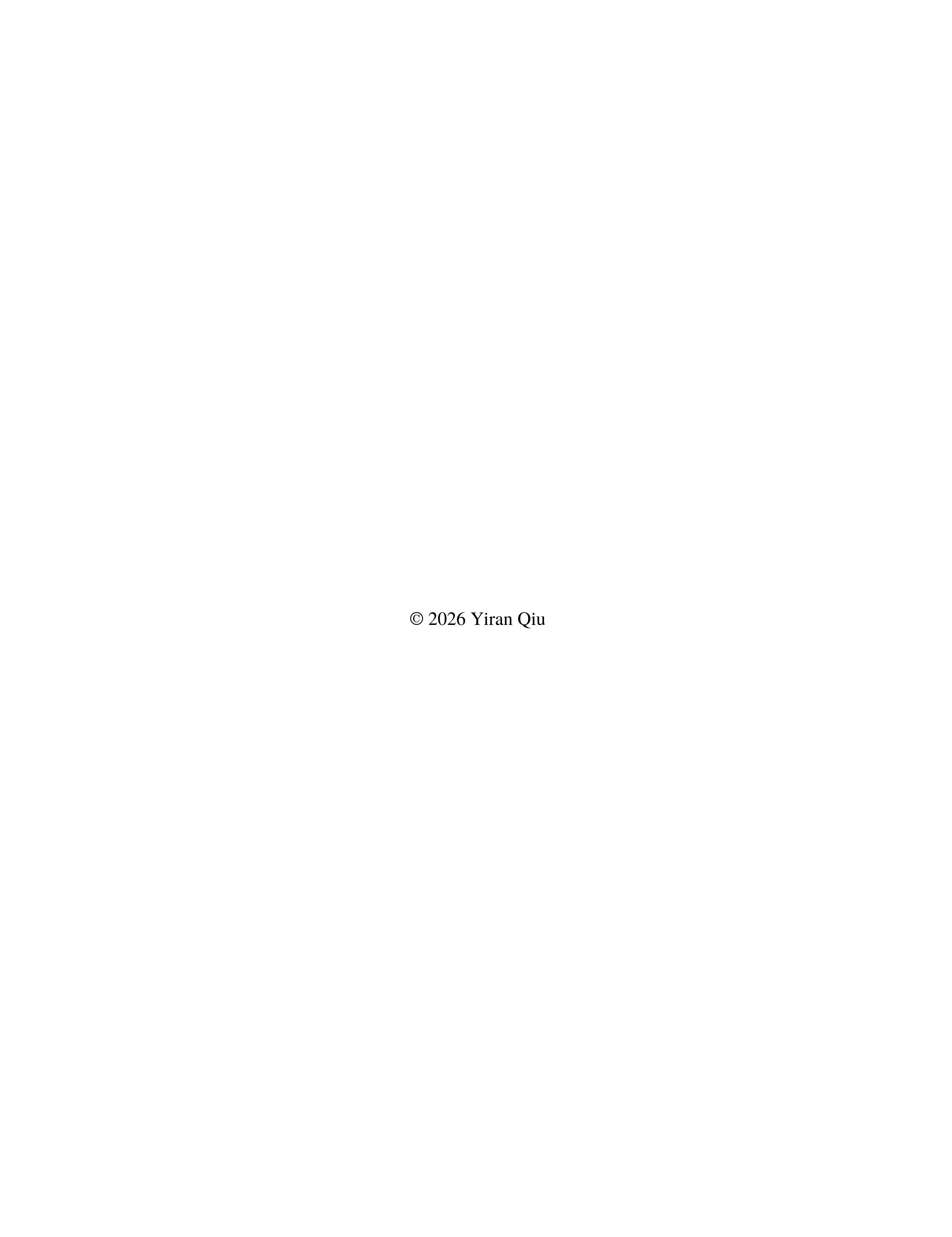

# OverrideFuzz: Semantic-Aware Grammar Fuzzing for Script-Runtime Vulnerabilities

BY

YIRAN QIU[1]

BACHELOR THESIS

Submitted in partial fulfillment of the requirements

for the degree of Bachelor of Science in Computer Science & Mathematics

in the College of Liberal Arts & Sciences

of the University of Illinois Urbana-Champaign, 2026

Urbana, Illinois

Advisor:

professor Nikita Borisov

[1] ORCID 0009-0004-0543-6011

# Abstract

Script-language runtimes such as Python, Lua, and JavaScript are widely deployed in security sensitive contexts, yet they remain difficult to test because valid inputs must satisfy syntax, dynamic type constraints, and object-level semantics. Existing grammar and reflection-based fuzzers improve syntactic validity and interface reachability, but they rarely model override hooks, dynamic rebinding, and attribute-resolution behavior that can redirect built-in operations across the script-native boundary and trigger use-after-free or type-confusion bugs.

We present OverrideFuzz, a two-phase, semantic-aware grammar fuzzer for script-language runtimes. Its declaration phase constructs objects with overriding methods, while its execution phase generates operations that route through those hooks. Active reflection tracks runtime types, and passive reflection learns from error messages to remove invalid operation shapes, allowing generation to approach semantic correctness without manual API specification.

We evaluate OverrideFuzz on CPython, Lua, and QuickJS. All three targets show consistent coverage growth, with rapid early expansion followed by slower incremental gains, and Lua benefits most from its pervasive metamethod dispatch mechanism. Although OverrideFuzz did not discover novel vulnerabilities during the bounded evaluation period, corpus analysis shows that it reconstructs inputs matching known vulnerability patterns, which suggests that semantic-aware generation reaches the intended script-native boundary behaviors.

*To my fellow squires,*

*may erudition light the way!*

# Acknowledgement

First, I would like to sincerely thank my advisor, Professor Nikita Borisov, for giving me the opportunity to develop this work into a thesis and for guiding me throughout the writing process. I am also deeply grateful to Professor Yan Shoshitaishvili from the SEFCOM Lab at Arizona State University, where this project began during the summers of 2024 and 2025. Without their support and guidance, I would not have been able to present this work here. Thanks for the trust and encouragement from the very beginning!

I would also like to thank the members of Ohio State Cybersecurity Club[1], SIGPwny[2], and R3kapig[3] for the many opportunities, support, and encouragement they have generously provided throughout my journey in cybersecurity. I am also grateful to my friends, Jack, Tom, Harley, Harry, Marshall, Conner, Antares, and everyone else for their companionship throughout my undergraduate studies. Beyond the technical communities that shaped this work, I would also like to thank the K-On and Vocaloid music community, whose creativity and music-making culture provided inspiration and balance throughout this work, 39!

Finally, I would like to thank my family for their unconditional love and support throughout my education, including the long journey that led to this thesis. Thanks for everything!

[1]Website: https://osucyber.club/
[2]Website: https://sigpwny.com/
[3]Website: https://r3kapig.com/

# TABLE OF CONTENT

# CHAPTER 1

# INTRODUCTION

How to test software in the development environment and find as many bugs as possible before release is one of the largest challenges in software security. A typical development cycle includes writing comprehensive unit tests. However, manually writing unit tests can require significant man-hours and is often difficult to make truly comprehensive. Therefore, techniques have emerged that trade man-hours for machine-hours.

For small software, one straightforward and robust approach is formal verification, which offers strong security guarantees and is commonly adopted in the aerospace industry under DO-178C [1]. For very simple programs, static analysis is often sufficient to cover all execution paths [2]. For slightly larger programs, a combination of static and dynamic analysis can be effective [3]. However, for extremely large codebases such as the Qt library, Chromium, or Linux, manual modeling becomes very difficult.

Therefore, it is important to adopt a testing method that can scale and efficiently trade man-hours for machine-hours. Fuzzing is an automated software testing technique that satisfies this requirement. In a nutshell, a fuzzer model can be defined as

$$\begin{aligned} m &: S \times H \times C \to H \\ e &: H \to C \end{aligned} \tag{1}$$

where $m$ is the mutation step and $e$ is the execution step. A *seed* $s \in S$ is an initial state of a randomness generator. A *harness state* $h \in H$ represents the fuzzer's generation and state at a given point, including the corpus of retained programs, the scheduler's counters, and any runtime context. A *coverage report* $c \in C$ is the set of control-flow edges exercised during execution. A control-flow edge is a branch transition between two consecutive basic blocks in the compiled target, recorded by lightweight compiler instrumentation, where a new edge indicates that a previously unexplored execution path has been reached. Given $m, e, s \in S, h_0 \in H$, the fuzzing process is defined by

$$
\begin{aligned}
h_{t+1} &= m(s, h_t, e(h_t)) \cup h_t \\
c_{t+1} &= e(h_t) \cup c_t
\end{aligned}
\tag{2}
$$

where the process keeps running until $\exists t, e(h_t)$ reports a crash.

## 1.1 Our contributions

Prior work has made substantial progress in generating valid programs and discovering callable runtime interfaces. However, existing approaches provide limited support for higher-level semantic features of dynamic languages, especially features that affect runtime dispatch and object interaction. This includes overriding, dynamic rebinding, and similar mechanisms that can serve as key primitives for vulnerabilities such as type confusion, memory corruption, and prototype- or object-state pollution.

To address the gap, in this paper, we discuss fuzzing applications for script-language runtimes and present OverrideFuzz, a platform for script-runtime fuzzing[4]. Our main contributions are:

- **Two-phase grammar fuzzing framework.** OverrideFuzz has a two-phase design that separates declaration-level generation (such as class definition, see Section 3.3.1) from execution-level generation (such as calling, see Section 3.3.2), enabling differentiated mutation over both program structure and runtime behavior (see Section 3.4).

- **Semantic-aware mutation strategy for dynamic languages.** We design mutators that explicitly exercise dynamic object semantics, such as overriding, rebinding, and attribute-resolution-related behaviors, in order to explore script-native boundary paths that are hard to reach with syntax-only mutations (see Section 3.3).

- **Practical implementation and empirical evaluation.** We implement OverrideFuzz as a reusable grammar-fuzzing pipeline (see Section 3) and evaluate it on representative script runtimes (CPython, Lua, QuickJS), analyzing corpus growth, and coverage effectiveness (see Section 4).

[4]Our fuzzer is open-source at github.com/Nambers/OverrideFuzz.

# CHAPTER 2

# BACKGROUND

## 2.1 Features

Script languages are widely adopted because of their flexibility and low programming burden. We begin by identifying the main characteristics of script languages to better target the fuzzing process.

**Dynamic Typing System.** Script-language interpreters do not fix variable types throughout their lifecycle. Instead, the interpreter binds variable names to objects, which provides high flexibility in programming. As a result, programmers no longer need enormous base classes with complex inheritance hierarchies, code generation, or templates just to satisfy a compiler. On the other hand, this feature also introduces substantial room for bugs such as type confusion, which can be triggered by passing unexpectedly typed objects as arguments, causing unexpected functions or attributes to be used.

**Reflection.** Although reflection does not exist only in script languages, it is more common and more useful there because of dynamic typing. Interpreters typically allow programmers to call built-in functions to inspect an object's type and other bound metadata. Therefore, designing a fuzzer that leverages reflection is important for obtaining the flexibility required to target the rich and diverse language features of script languages.

**Rich Built-in Modules.** Most script-language interpreters are shipped with a bundle of built-in modules, similar to a standard library. However, many of these modules are not written in the script language itself. Such an interpreter is said to be not fully bootstrapped: its core is implemented in a lower-level language such as C rather than in the script language it executes. This design extends functionality but also introduces additional vulnerabilities. These interpreter components are hidden behind the language parser and are often hard to reach without significant manual effort in writing code generators. By leveraging reflection, we can dynamically discover function and variable properties under modules or classes, giving the fuzzer additional capability to discover valid paths.

## 2.2 Related works

**Table I**

**Comparison of fuzzing approaches**

| Fuzzer | Input Representation | Coverage-guided | No seed corpus | Reflection assisted | Override semantics | Multiple runtime supported |
|---|---|---|---|---|---|---|
| AFL++ | Text mutation | ✓✗ | × | × | ✓✗[5] | ✓ |
| Superion | Tree mutation | ✓ | × | × | ✓✗[5] | ✓ |
| Nautilus | CFG[6] gen. + mut. | ✓ | ✓ | × | × | ✓ |
| PolyGlot | IR mutation | ✓ | × | × | ✓✗[5] | ✓ |
| Fuzzilli | IR gen. + mut. | ✓ | ✓ | × | ✓✗[5] | × |
| PatchFuzz | Text mutation | ✓ | × | × | ✓✗[5] | × |
| Reflecta | CFG gen. + mut. | ✓ | ✓ | ✓ | × | ✓ |
| OverrideFuzz | AST gen. + mut. | ✓ | ✓ | ✓ | ✓ | ✓ |

Existing fuzzing techniques for script-language runtimes mainly fall into four categories: syntax-aware program generation, reflection-assisted exploration, general coverage-guided runtime fuzzing, and corpus-driven mutation (see Table I). These directions improve the validity and reachability of generated programs, but they do not fully model the higher-level object semantics commonly used in dynamic languages.

**IR-based and grammar-based fuzzing.** Fuzzilli [4] and Nautilus [5] (with ANTLR v4 grammar IR) present advanced IR designs and conduct IR-based fuzzing, offering strong flexibility across different script languages. PolyGlot [6] is another work that mutates IR with constrained mutation kinds to preserve a high grammar-correctness rate without losing too much discovery capability. These approaches are well suited for driving execution into parsers, bytecode compilers, and general runtime components. However, their primary focus is structural validity and generic program mutation. They do not explicitly model object-level semantic relations such as method overriding, dynamic rebinding, custom attribute resolution, or other dispatch-affecting behaviors.

[5]corpus-dependent: possible only if seed corpus contains override patterns
[6]context-free grammar

As a result, they may miss bugs whose triggering conditions depend on carefully constructed runtime object semantics rather than syntax alone.

**Corpus-driven mutation.** Some fuzzing approaches rely on an existing corpus of inputs as the primary material for generating new test cases, either by mutating seeds drawn from the corpus or by combining fragments across corpus entries. PatchFuzz [7] exemplifies this strategy in the context of JavaScript engine fuzzing. It automatically extracts proof-of-concept inputs from historical security-fix commits, uses those inputs as seeds, and applies selective instrumentation to concentrate fuzzing resources on code regions modified by the corresponding patches. This design enables efficient rediscovery of vulnerability families near known bug sites. Similarly, combination mutations, such as the combine mutation in Fuzzilli [4], which inserts a complete IR program drawn from the retained corpus into another at a random position, depend on the same principle: richer and more diverse corpora yield more varied combinations and broader coverage. The shared limitation across these approaches is that the effective discovery space is bounded by the patterns already present in the corpus. If no retained entry contains an override-bearing subclass or a custom attribute-resolution hook, no combination or mutation of existing programs can introduce that construct.

**Reflection-based fuzzing.** Reflecta [8] develops a fuzzer that leverages reflection capabilities. However, among these works, to the best of our knowledge, few explicitly utilize higher-level script-language features such as overriding, which is an important primitive for prototype pollution. Prototype pollution is a class of vulnerabilities in which attacker-controlled code modifies a shared base-class or prototype object so that all instances derived from it unexpectedly inherit implanted malicious properties or behaviors. This direction is particularly relevant to script languages, where rich introspection support is often available by design. Still, reflection mainly solves interface discovery. It does not by itself provide a mechanism for constructing semantically meaningful object states or interaction patterns. In particular, knowing that an object is callable or that a method exists is insufficient for generating inputs that alter dispatch behavior through overriding or related dynamic language features.

**Coverage-guided fuzzing for runtimes and built-in components.** Coverage-guided fuzzers track which control-flow edges are exercised by each generated input and prioritize inputs that reach previously unseen edges. General coverage-guided fuzzing, such as AFL++ [9], remains the dominant strategy for large-scale software testing. It is effective for exploring large codebases with minimal manual modeling, and it can be combined with custom grammar-based generation or reflection-based mutator modules, for example, Superion [10] implements tree-level mutation under AFL++. For script-language runtimes, however, many vulnerable paths lie behind semantic checks at the script-native boundary. Reaching such paths often requires more than valid syntax or broad API enumeration: the fuzzer must construct objects whose runtime behavior violates implementation assumptions. Examples include user-defined objects that override built-in behavior with side-effect functions, customize attribute access, or satisfy a protocol in an unexpected way. These conditions are difficult to infer through generic mutation alone.

**Preliminary work.** Prior to OverrideFuzz, we developed PyFuzzer [11], a CPython-specific prototype that mutated Python's native abstract syntax tree (AST) directly and executed generated programs via CPython's internal `run_mod` interface, instrumented with libFuzzer and AddressSanitizer. This approach was abandoned for two reasons. First, constructing mutations at the AST level can skip the parser and proceed directly into the interpreter, but it required calling CPython's specific node-constructor functions with precisely typed arguments for every node kind, resulting in high manual engineering overhead for each new mutator and large amounts of helper code. Second, CPython alone is not a high-value target in the threat model we care about: Python is rarely deployed as a sandboxed or embedded scripting layer with a restricted execution surface, so vulnerabilities found there have limited practical impact compared to runtimes such as Lua or QuickJS, which are routinely embedded inside host applications with a clear trust boundary. These two limitations together motivated the language-portable (see Section 3.6), grammar-level design of OverrideFuzz.

## 2.3 Threat Model

**Attacker position.** We consider an attacker who controls the script input submitted to a deployed interpreter. In the game industry, this corresponds to a player or modder who can supply Lua scripts that the host application executes with minimal auditing. In browser-adjacent or enterprise deployment contexts, this corresponds to an untrusted user whose JavaScript or Python code is evaluated by an embedded runtime behind a restricted API surface. The attacker cannot modify the interpreter binary or the host application, nor invoke native system APIs directly. The sole capability the attacker possesses is the ability to submit script code.

**Trust boundary.** The interpreter enforces a boundary between script-managed and host-managed execution. Script code is confined to the runtime's object model, and the host process, including the interpreter's own C implementation, lies outside this boundary. A correct interpreter guarantees that script code cannot access arbitrary host memory, corrupt the runtime's internal state, or transfer execution outside the permitted interface.

**Vulnerability class.** This boundary can be violated when the interpreter's C implementation fails to account for user-defined hooks that execute during built-in operations. When the interpreter invokes a user-defined method as part of an operation such as indexing, comparison, or attribute resolution, it temporarily yields control back to the script layer. If the C implementation holds a live pointer to an object during this period, and the user-defined hook frees or reallocates that object as a side effect, the interpreter resumes with a dangling pointer. The resulting use-after-free or type-confusion condition constitutes a memory-safety violation that can allow the attacker to read or write arbitrary host memory, thereby escaping the sandbox.

**Attacker goal.** The attacker seeks a sandbox escape: the ability to execute arbitrary code in the host process. We model this concretely as triggering a memory-safety violation within the interpreter's C implementation that AddressSanitizer detects. Reaching such a primitive is sufficient to demonstrate the presence of the vulnerability. Full exploit development is outside the scope of this work.

**Scope.** The threat model covers script-language runtimes that are deployed with a restricted API surface intended to isolate the script layer from the host and that implement built-in types and operations in a lower-level language such as C. CPython, Lua, and QuickJS each satisfy both conditions and were selected accordingly (see Section 4.1). Vulnerabilities arising in host application code beyond the interpreter boundary, in transport or network layers, or in the script-language specification itself are out of scope.

## 2.4 Motivation Examples

A *Use-After-Free* (UAF) is a memory safety vulnerability in which a program continues to access memory after the allocator has reclaimed it. In managed language runtimes, UAFs typically arise when user-defined override hooks trigger unexpected deallocation during an operation the interpreter assumes to be safe, allowing the freed region to be reallocated and corrupted by subsequent code. Both examples below were detected with AddressSanitizer (ASan), a compiler-instrumented runtime detector that intercepts illegal memory accesses and reports them as errors.

The following examples are drawn from CPython issue reports and serve as concrete instances of a broader vulnerability class that motivates the design of OverrideFuzz, namely bugs triggered by user-defined objects that override built-in dispatch hooks with unexpected side effects. Although both examples are presented in Python, this vulnerability class is not unique to CPython, as equivalent triggering conditions arise in other script runtimes through analogous override mechanisms, including Lua table metamethods and QuickJS Proxy or prototype-chain traps. These cross-runtime mechanisms are precisely what OverrideFuzz is designed to exercise, and the CPython examples below are therefore representative of a general pattern rather than a platform-specific concern.

In the Python snippet below (see Example I) [12], CPython only checks whether the length remains unchanged, but accidentally holds a pointer to freed memory. Therefore, after freeing the old object and creating a new one in place, then restoring the length attribute, we can write bytes directly into the new object's memory region. After writing a `1` at the 23rd position to tweak the length, we obtain an arbitrary read/write primitive.

```
class UAF:
        def __index__(self):
                global memory
                uaf.clear() # uaf is in captured closure
                memory = bytearray() # use-after-free
                uaf.extend([0] * 56) # bypass the length check
                return 1

uaf = bytearray(56)
uaf[23] = UAF()
# memory is UAF right now, UAF reported by ASan

# -- exploit --

# memory[id(20) + 24] = 114
# print(20) # = 114
```

**Example I. A CPython bytearray UAF**

In the Python snippet below (see Example II) [13], CPython fails to check whether the passed argument is modifiable, which leads to accidental freeing of a built-in buffer and replacement with a custom object. This eventually yields a use-after-free primitive that can enable type object corruption.

```python
class A:
        def __eq__(self, other: dict):
                # expressions below are equivalent

                # del other["items"]

                # this will free the original function
                # by `decref` in `insertdict` function
                other["items"] = 0
                # other["items"] = []
                # other["items"] = bytearray(100)

# it will modify the member function in dict, and it should not?
A().__eq__(dict.__dict__) # UAF reported by ASan
# {1:1}.items() # crash

# -- exploit --
# print(dict.items) # print will fill itself into `dict.items` freed buffer
# dict.items("output\n") # equivalent to `__import__("sys").stdout.write`
# dir(dict.items) # or dir will fill itself too
# dict.items == dir
```

**Example II. A CPython type confusion UAF**

# CHAPTER 3

# IMPLEMENTATION

## 3.1 Internal AST

An *Abstract Syntax Tree* (AST) is a tree-structured representation of a program where every node corresponds to a syntactic construct such as a function definition, an operator expression, or a variable declaration, independent of any target language's concrete syntax. OverrideFuzz generates and mutates programs entirely as ASTs. Without losing the ability to target different languages, the mutation engine operates on an internal AST schema with universal rules for most languages, listed below using Zephyr ASDL notation (see Schema I) [14]. The schema is a subset of the grammar of the target languages, which allows the serializer to produce valid parser inputs directly from the tree.

```
-- ASDL's 4 builtin types are:
-- identifier, int, string, constant

binary_op    = Add | Sub | Mul | Div | Mod | Pow | Eq | NotEq | Lt | Gt | LtE |
GtE | BitAnd | BitOr | BitXor | LShift | RShift
unary_op     = Neg | Not | BitNot

function     = Function(identifier name, identifier* args, scope scope)
class        = Class(identifier name, class* cases, function* funcs)
declareVar   = DeclareVar(identifier name, constant val)
import       = Import(identifier lib_name)

declaration_stmt  = function | class | declareVar | import
declaration_scope = Scope(declaration_stmt* stmts)

object       = identifier | call

getProp      = GetProp(identifier l, identifier r, object attr)
setProp      = SetProp(identifier l, object r, object attr)
call         = Call(identifier name, object* args)
return       = Return(object? val)
```

```
binaryOp     = BinaryOp(identifier l, identifier a, binary_op op, identifier b)
unaryOp      = UnaryOp(identifier l, unary_op op, identifier r)
newInstance = NewInstance(identifier l, identifier class_name, object* args)
getItem      = GetItem(identifier l, identifier r, object idx)
setItem      = SetItem(identifier l, object idx, object r)

execution_stmt  = getProp | setProp | call | return | binaryOp | unaryOp |
newInstance | getItem | setItem
execution_scope = Scope(execution_stmt* stmts)

globalRef = GlobalRef(identifier* refs)
scope     = Scope(globalRef ref, declaration_scope declaration, execution_scope
execution)
```

**Schema I. Internal AST**

## 3.2 Configuration parameters

**Table II**

$w_{\text{decl}}$ **Sampling weights for declaration mutators (left), and** $w_{\text{exec}}$ **execution mutators (right).**

| Operation | Weight | % |
|---|---|---|
| AddFunction | 30 | 44.1 |
| AddVariable | 20 | 29.4 |
| AddClass | 12 | 17.6 |
| AddImport | 6 | 8.8 |
| **Total** | 68 | 100 |

| Operation | Weight | % |
|---|---|---|
| GetItem | 15 | 18.1 |
| SetItem | 15 | 18.1 |
| Call | 14 | 16.9 |
| SetProp | 13 | 15.7 |
| GetProp | 12 | 14.5 |
| NewInstance | 9 | 10.8 |
| BinaryOp | 2 | 2.4 |
| Return | 2 | 2.4 |
| UnaryOp | 1 | 1.2 |
| **Total** | 83 | 100 |

**Table III**

**Configuration parameters used by OverrideFuzz.**

| Symbol | Default Value | Description |
|---|---|---|
| $t_{\text{floor}}$ | 100 | Minimum trials for execution generation without coverage |
| $t_{\text{ceil}}$ | 2000 | Maximum trials for execution generation without new coverage |
| $s_{\text{cap}}$ | 50 | Maximum scope count in declaration mutation |
| $w_{\text{decl}}, w_{\text{exec}}$ | see Table II | Sampling weights for mutators |
| $w_{\text{startover}}$ | 10% | Weight of startover from empty corpus |
| $w_{\text{respectType}}$ | 80% | Weight of respecting type hint |

| Symbol | Default Value | Description |
|---|---|---|
| $w_{\text{var}}$ | $(9, 1, 6)$ | Weight of feeding non-const, const and function call as parameter |
| $t_{\text{line}}$ | $500ms$ | Execute one line timeout |
| $t_{\text{lines}}$ | $1000ms$ | Replay multiple lines timeout |

## 3.3 Mutators: Input Generation

There are two groups of mutators, designed for declaration statements and execution statements. In general, we can classify most script-language code lines into these two categories. For conciseness, we treat class definition, function definition, import statement, and variable statements as declarations because they introduce new bindings into the context. Then we treat operators, call patterns, and property/item accesses as executions because they change the state of existing bindings.

Both phases communicate through a shared binding context, a symbol table mapping each identifier to its kind (class, function, variable, or imported module). The execution mutator uses this table as its operand pool after declaration mutation completes, and reflection results augment it at runtime with names that were not statically visible at generation time. Nested scopes additionally rely on the `globalRef` field of each `Scope` node, which lists the outer identifiers that an inner function body captures. The generator later populates this field so that override bodies can reference and mutate objects declared in enclosing scopes.

**Declaration Mutators.** The declaration mutator extends the `declaration_scope` of a given scope node and is responsible for constructing the symbolic context on which execution statements subsequently operate. OverrideFuzz supports four mutation operations for declaration.

1) `AddClass` appends a `Class` node inheriting from a base class drawn from the built-in or library types. It provides the structural complexity that OverrideFuzz needs to reach non-trivial paths.

2) `AddFunction` appends an overriding `Function` node under an existing `Class` node. The target class is randomly drawn from all custom class nodes in the AST. Function signatures are drawn from reflection results on the base class of the selected class node. The mutator then recurses by first applying declaration mutation to the new function scope and then execution mutation, so that override bodies can themselves contain complex operations. This is the primary mechanism by which OverrideFuzz generates objects with non-trivial runtime semantics rather than inert data containers.

3) `AddVariable` appends a `DeclareVar` node with a type drawn from basic types of the language. If its type is numerical, like integer, boolean, the values are drawn from structured random generation. If its type is buffer-like, values are produced by *havoc*, a strategy of applying random bit-level perturbations to a byte buffer, to probe edge cases in interpreter value handling. The custom class type values are generated by the execution mutation.

4) `AddImport` appends an `Import` node for a module selected from the target's built-in registry. Importing a module introduces its constants, classes and member functions available to be overridden or called, extending the mutator's reach to the built-in modules of the interpreter.

Mutation kind selection is weighted toward `AddFunction` and `AddClass`, which introduce the semantic complexity on which execution mutations most depend, with `AddVariable` and `AddImport` receiving lower weight (see $w_{\text{decl}}$ in Table II).

**Execution Mutators.** While additional mutation operations are conceivable, for example control flow mutation like inserting loop constructs or conditional branches, these tend not to contribute meaningfully to vulnerability discovery and instead degrade throughput by stalling execution, producing unreachable code, or disrupting control flow. OverrideFuzz therefore focuses on six kinds of execution mutation, comprising nine operations in total, each chosen for its ability to drive execution into semantically interesting interpreter paths.

1) `GetItem` and `SetItem` append index-based read and write expressions onto the execution scope. Python and similar dynamic languages allow user-defined classes to customize built-in operations through *dunder* (double-underscore) methods: specially named methods such as `__getitem__` / `__setitem__` (proxy `[]` operator), and `__index__` (determines the actual index of the class) that the interpreter calls instead of its built-in implementation when an operation is applied to a user object. Beyond simple buffer access, these operations are particularly effective when the target object carries such overridden methods implemented with unexpected side effects, as the interpreter must then dispatch through the user-supplied hook as illustrated in Example I.

2) `Call` appends an invocation of a callable drawn from the reflection-populated context, allowing greater variable variety. Argument types are constrained by live binding information collected during execution, which increases the proportion of well-typed calls and reduces trivial type errors that would terminate execution before reaching deeper interpreter paths.

3) `Return` appends an exit statement at the end within the enclosing function scope, providing the conventional return-value mechanism for data exchange beyond side effects.

4) `UnaryOp` and `BinaryOp` append operator expressions whose operands are drawn from active context bindings. For user-defined objects, operator application triggers dunder dispatch, such as __`add`__, __`eq`__, or __`lt`__, creating a direct path from the override mutations introduced in the declaration phase into interpreter arithmetic and comparison routines.

5) `NewInstance` appends an instantiation of a custom class, introducing freshly typed objects into the execution context for subsequent mutations to operate on.

6) `GetProp` and `SetProp` append attribute reads and writes. When the target object carries an overridden __`getattr`__ or __`setattr`__, these operations enter the attribute-resolution machinery of the interpreter and can expose bugs related to prototype-state pollution or unexpected freeing of internal attribute buffers, as illustrated in Example II.

Mutation kind selection among execution operations is weighted toward `GetProp`, `SetProp`, `Call`, `GetItem`, and `SetItem`, as these operations directly trigger dispatch through user-defined attribute and index hooks, which are the interpreter paths most exposed by the override mutations introduced in the declaration phase (see $w_{\text{exec}}$ in Table II). `BinaryOp` and `UnaryOp` receive moderate weight because operator dunder dispatch is a frequent source of type-confusion paths. `Return` and `NewInstance` receive the lowest weight, as they contribute structural scaffolding rather than dispatch diversity.

## 3.4 Scheduler: Generation Guide

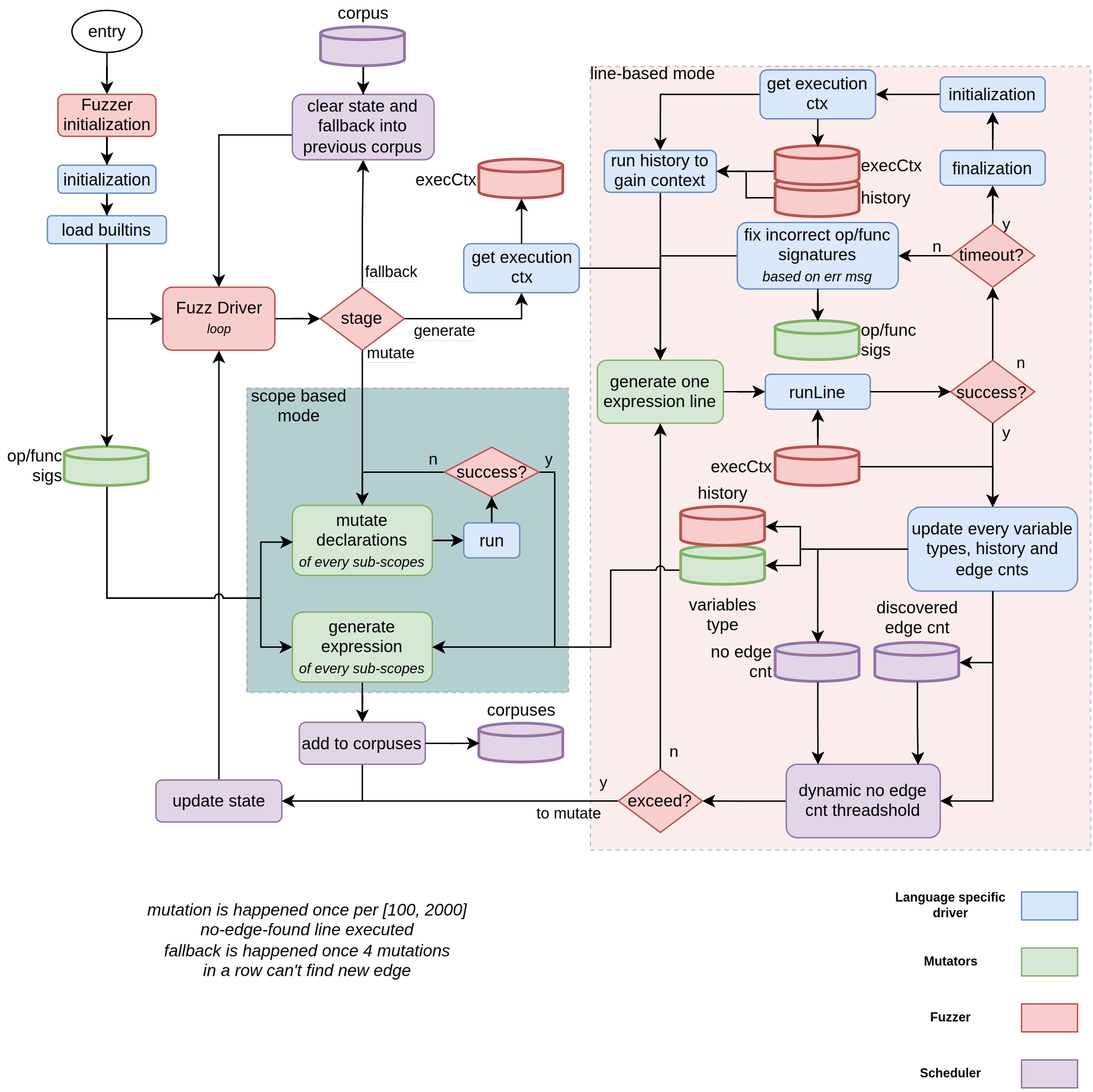

**Fig. 1. Architecture of OverrideFuzz**

Fig. 1 illustrates the overall generation pipeline of OverrideFuzz, which includes dispatching, mutating, and coverage feedback collection.

**Dispatch.** The *corpus* is the set of programs the fuzzer retains as seeds for future mutation. A program is added to the corpus only when its execution discovers at least one new coverage edge, ensuring that every retained entry expands the explored space. Unlike mutation-based fuzzers that usually require a user-supplied seed corpus to bootstrap exploration, OverrideFuzz starts from a single blank AST and grows its corpus entirely through its own generation. On initialization, if no prior session is loaded, a minimal program is synthesized via the declaration mutators to serve as the first corpus entry. Subsequent corpus entries are accumulated only when a generated program discovers new coverage edges. When execution generation stalls, the FallbackOldCorpus phase selects a previously discovered entry at random and resumes mutation from that state, occasionally restarting from the blank entry with $w_{\text{startover}}$ probability to prevent the search from converging prematurely. This design removes the curation burden of assembling representative seed inputs for each new interpreter target.

**Mutation scale.** OverrideFuzz applies two distinct mutation granularities depending on whether the current stage targets declarations or execution expressions, as shown in the green *Mutators* nodes of Fig. 1.

Declaration mutations operate in scope-based mode, where all sub-scopes are processed together in a single pass. Because declarations are lazy-executed, their effect is only observable when later execution code invokes them, so finer per-line control yields no additional feedback or benefit, while it also introduces problems further discussed in Section 6.1. To avoid thrashing the declaration scope while execution is still discovering coverage with the existing definitions, a declaration mutation is triggered only once per dynamic window of between $t_{\text{floor}}$ and $t_{\text{ceil}}$ consecutive no-edge lines executed (see Table III). The window adapts to the rate at which execution finds new coverage so that declaration mutations remain infrequent when execution is productive.

Execution expressions use line-based mode, where a single expression line is generated, immediately dispatched to the language-specific driver, and its result is observed before the next line is constructed. This granularity is essential for reflection, as the outcome of each line, which may be a new coverage edge or a runtime error, is fed back to the scheduler and mutators before the next step is taken, allowing per-line correction of operator and function signatures (see Section 3.5).

**Coverage feedback.** After each `runLine`, the purple *Scheduler* nodes in Fig. 1 evaluate the execution outcome along two dimensions: coverage progress and error signal.

After the line executes, the scheduler updates the variable type map, the operation history, and the per-edge counter for the current corpus entry. New edges are detected via SanitizerCoverage [15] (see Snippet II). When no new edge is found, the no-edge counter increments. Once the counter exceeds the dynamic threshold, the scheduler triggers a declaration mutation pass to introduce new callable hooks and class definitions that may open unexplored dispatch paths. The threshold is computed as

$$T_{\text{stall}} = \min(\max(\lfloor \log_2(N + 4) \times 50 \rfloor, t_{\text{floor}}), t_{\text{ceil}}) \tag{3}$$

where $N$ is the total number of new edges discovered across the entire run. Early in the run, when $N$ is small, the threshold evaluates to $t_{\text{floor}}$, so declaration mutations fire frequently to help bootstrap coverage. As $N$ grows, the threshold increases logarithmically toward $t_{\text{ceil}}$, allowing execution to explore longer before re-mutating declarations. If the current scope count has already reached the cap of $s_{\text{cap}}$ scopes, the scheduler skips declaration mutation and enters fallback directly, since adding further declarations would not expand the reachable dispatch surface. If five consecutive declaration mutation rounds each fail to produce a new edge, the fallback phase activates. The scheduler clears the current generation state and restores a randomly selected prior corpus entry, or reverts to the blank entry with 10% probability, before resuming generation, in order to prevent the search from converging on a narrow region of the coverage space.

Runtime errors returned by the driver are not discarded. Instead, they are forwarded to the passive reflection layer, which parses the error message to correct operator signatures and function call shapes in the current context memory. The corrected signatures are applied starting from the next generated line, reducing the rate of grammatically invalid programs in subsequent execution steps.

**Timeout Recovery.** Because line-based execution is stateful, each generated line may reference variables, class instances, or function definitions introduced by earlier lines in the same session. When a `runLine` call in the execution generation phase exceeds the per-line budget $t_{\text{line}}$, the interpreter is killed and restarted by `longjmp`, and the accumulated execution context, including all variable bindings and class hierarchies built up to that point, is corrupted. To avoid discarding the entire session, the scheduler maintains a history of all lines that completed successfully before the timeout. After restarting the interpreter, OverrideFuzz replays those lines against the fresh process, subject to the aggregate replay budget $t_{\text{lines}}$. Because replay re-executes only previously validated lines, it is expected to complete well within $t_{\text{lines}}$ under normal conditions. Once replay finishes, the scheduler resumes generation from the point immediately following the timed-out line, preserving the semantic context that was accumulated before the interruption.

## 3.5 Reflection: Semantic Correctness

Because dynamic languages do not enforce type constraints at definition time, the fuzzer cannot determine at generation time whether an operation on a variable will succeed at runtime. OverrideFuzz addresses this through two complementary reflection strategies: active reflection, which queries runtime state before or after each generation step, and passive reflection, which updates the context model in response to runtime errors.

**Active reflection.** OverrideFuzz performs two kinds of active reflection. First, after each successful `runLine`, it queries the live execution context to read the concrete runtime type of every in-scope variable using the interpreter's own type introspection API (for example, `Py_TYPE` in CPython). The resolved type is written back to the AST variable record so that subsequent mutations select operations that are valid for the actual runtime type rather than a stale inferred one. Second, at startup OverrideFuzz performs an initial scan of all known built-in types and modules to populate the set of overrideable function signatures available to `AddFunction`. For each built-in class or module (for example, `bytearray` and `list` in CPython, table metamethods in Lua, and built-in constructors in QuickJS), OverrideFuzz inspects available override entry points through the interpreter's own introspection API. Representative examples of functions discovered through this process are listed in Table V.

**Passive reflection.** When `runLine` returns an error, OverrideFuzz parses the error message to identify which combination of operation and type was invalid, then applies one of four targeted corrections to the shared context model. OverrideFuzz recognizes four classes of runtime error and applies a corresponding targeted correction to the shared context model for each: removing an incompatible type from an operator's allowed type set, evicting a missing attribute from a type's property list, updating a parameter type at the faulting position, or correcting the stored arity of a function. Representative error messages and the correction each triggers are listed in Table VI. Each correction is permanent for the remainder of the session. The same invalid combination is therefore never generated again, progressively tightening the generation space toward semantically valid programs without requiring any manual specification of the target API.

**Type hint violation cases.** Script languages have weak typing, where a function that declares a parameter as `int` will generally accept any value without a static check, and the interpreter resolves the mismatch at the call site rather than at parse time. OverrideFuzz exploits this property deliber-ately. When generating a function call, the argument type drawn from the active reflection context is respected with probability $w_{\text{respectType}}$. For the remaining calls, OverrideFuzz intentionally supplies a value whose runtime type differs from the recorded type hint. This controlled mismatch is a direct attempt to explore type-confusion bugs. If the callee's implementation assumes the argument has a particular metamethod and does not validate that assumption, feeding an unexpected type can call to a different method without raising error, and corrupt the presumption and internal state.

## 3.6 Language Portability

The mutation engine, scheduler, and internal AST described in the preceding sections are entirely language-agnostic. All language-specific logic is encapsulated to a per-target driver, represented as the blue nodes in Fig. 1. A driver implements four responsibilities: serialization, execution, active reflection, and passive reflection.

**Serializer.** The serializer translates each internal AST node into an executable surface representation for the target language. For constructs that exist directly in the target language, this translation is mostly syntactic. For example, a `Call(f, args)` node becomes a function invocation in CPython, Lua, or QuickJS. For constructs whose surface forms differ across languages, such as class construction and override dispatch, the driver lowers the same internal node into the target's object mechanism. A `NewInstance(l, C, args)` node therefore serializes to ordinary class instantiation in CPython, to a metatable-backed constructor pattern in Lua, and to constructor invocation in QuickJS. This design preserves a shared mutation interface while allowing each driver to encode the runtime-specific mechanisms needed to exercise override hooks.

**Execution driver.** The execution driver embeds or invokes the interpreter and implements the `runLine` interface: it accepts a serialized line, submits it to the interpreter, and returns either a coverage delta or an error message string. Coverage instrumentation differs per target. CPython and QuickJS are instrumented with SanitizerCoverage edge callbacks compiled in at build time. Lua uses a custom hook registered through the Lua C API that records branch transitions at the C level of the interpreter.

**Active reflection adapter.** The active reflection adapter implements the two runtime queries described in Section 3.5. The per-line type read-back uses `Py_TYPE` in CPython, `lua_typename` together with the debug library in Lua, and `typeof` together with `Object.getPrototypeOf` in QuickJS. The startup built-in scan enumerates overrideable entry points using `dir()` on built-in classes in CPython, the `__index` metamethod table for standard types in Lua, and `Reflect.ownKeys` on built-in constructors in QuickJS.

**Passive reflection parser.** Because each interpreter formats runtime error messages differently, the passive reflection parser is also target-specific. The parser consists of a small set of regular expressions that map interpreter error strings to one of the four correction types. The correction types themselves are shared across all targets, but the patterns that recognize them differ per target. Representative error strings and their corresponding corrections are listed in Table VI.

Therefore, adding a new target requires implementing a serializer, a `runLine` harness with appropriate coverage instrumentation, an active reflection adapter, and a set of error-pattern regexes for passive reflection. The other components described above need no changes.

## 3.7 Reproducibility

OverrideFuzz uses Nix Shell to provide a fully reproducible build and execution environment [16]. Each target builds its interpreter from source inside a Nix derivation that pins the exact versions for interpreter source archive, compiler, linker, and all library dependencies.

# CHAPTER 4

# EVALUATION

## 4.1 Targets and Experiment Setup

We evaluate OverrideFuzz against three scripting runtimes: CPython version 3.14.3 [17], Lua version 5.5.0 [18], and QuickJS version 2025-09-13 [19], each being the latest available version when this project began, around late 2025. These targets were selected to match the threat model (see Section 2.3). CPython is the reference implementation of Python and the most widely deployed scripting language runtime in production systems. Lua is the dominant embedded scripting engine in the game industry, where it mediates a high-frequency interface between native C/C++ hosts and script logic. QuickJS is a compact, self-contained JavaScript engine that implements a recent ECMAScript specification [20] and represents the class of lightweight JS runtimes critical to browser-adjacent and embedded contexts.

All experiments were conducted on a single machine running Arch Linux (kernel `6.19.11-arch1-1`), equipped with an AMD Ryzen 9 9900X processor (12 cores, 24 threads) and 32 GB of RAM. Each fuzzer instance runs on a single thread without parallelism.

## 4.2 Research Questions

In evaluation, we mainly measure performance based on the following research questions.

**RQ1.** How much source-level coverage does OverrideFuzz achieve across targets? (Section 4.3)

**RQ2.** Does OverrideFuzz discover novel vulnerabilities within the bounded evaluation campaign? (Section 4.4)

## 4.3 Coverage

To answer RQ1, OverrideFuzz achieves consistent and measurable coverage growth across all three target runtimes. Fig. 2, Fig. 3, and Fig. 4 plot function, line, and branch coverage over the course of each campaign, and Fig. 5 through Fig. 7 show the per-file coverage distribution at campaign end.

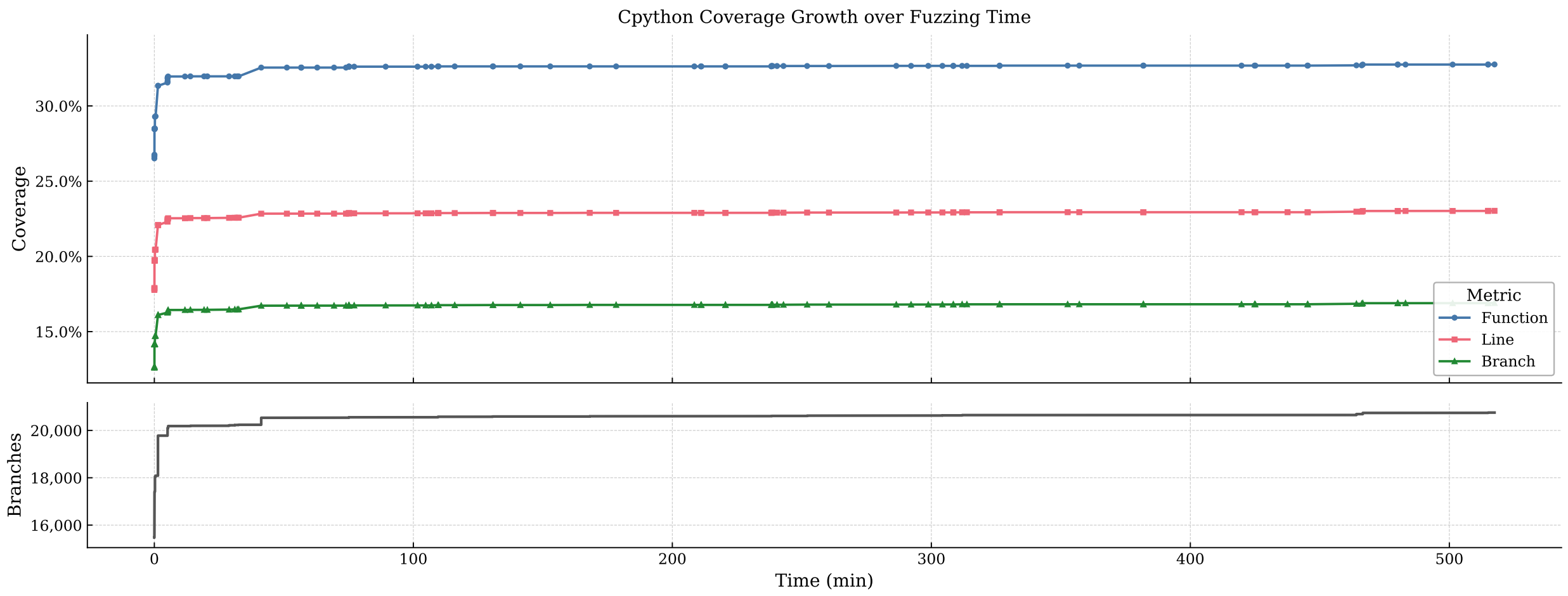

**Fig. 2. CPython coverage growth over time**

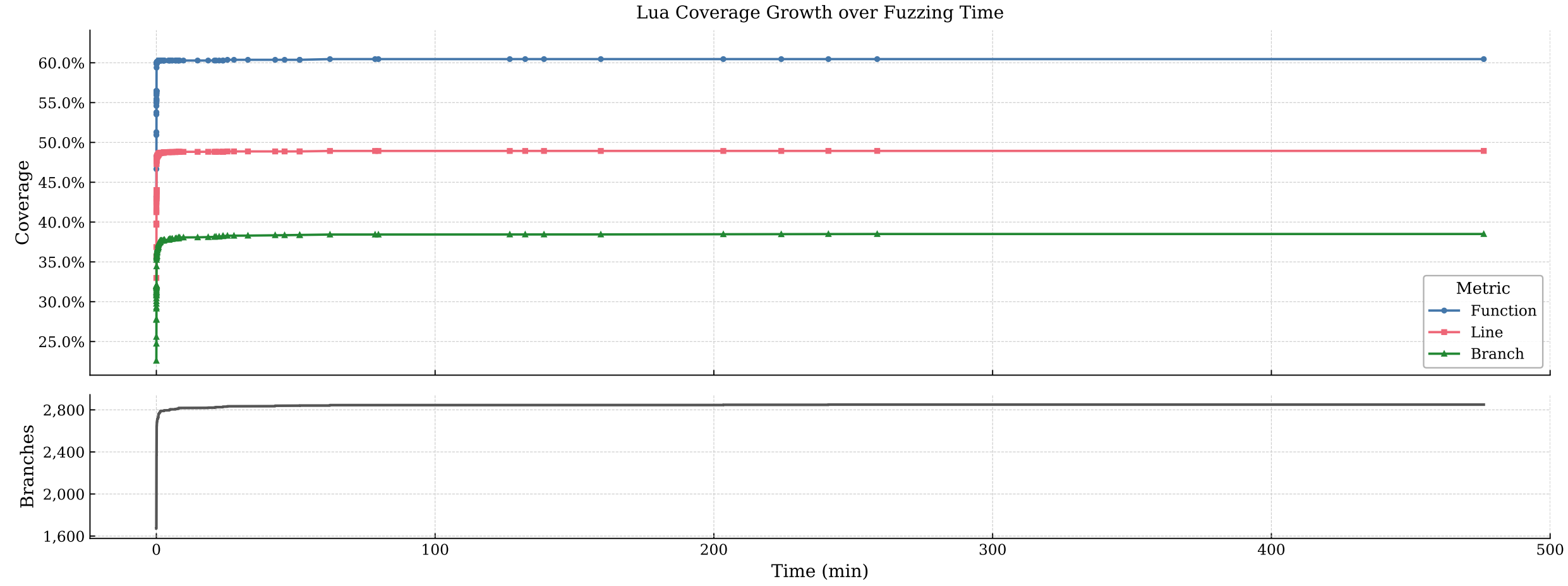

**Fig. 3. Lua coverage growth over time**

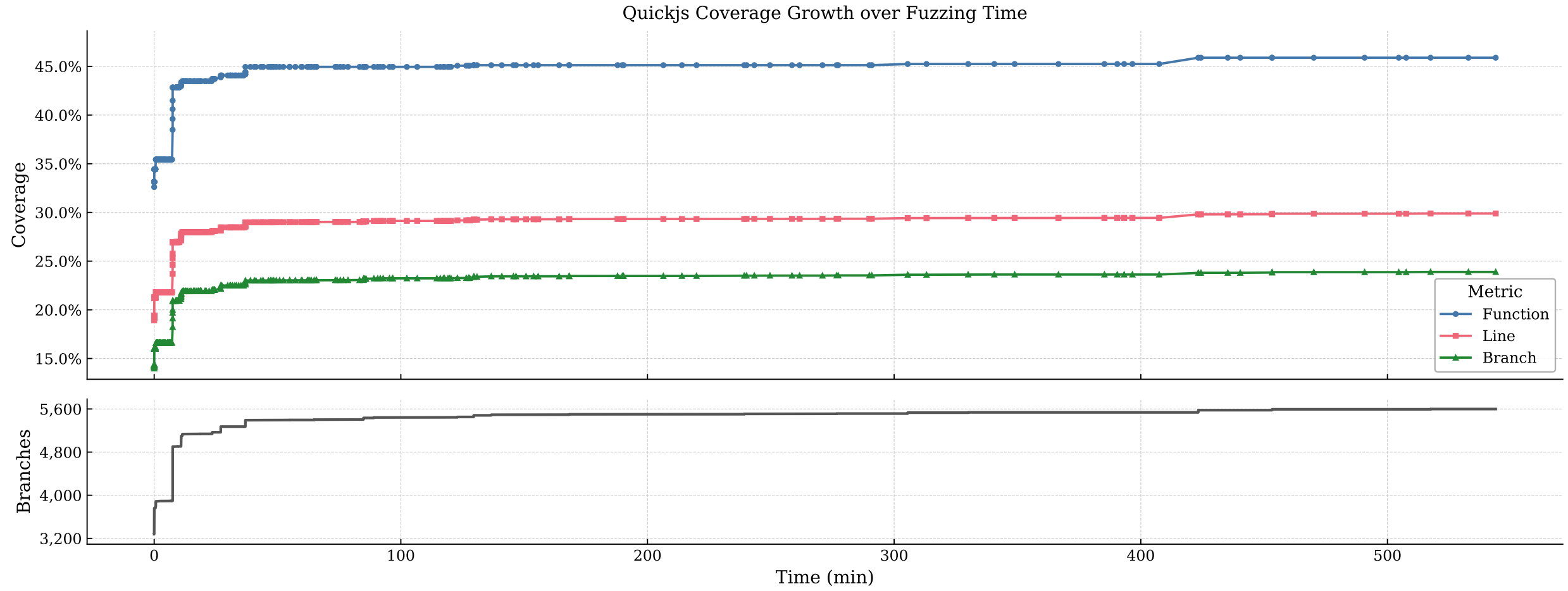

**Fig. 4. QuickJS coverage growth over time**

All three runtimes exhibit a characteristic two-phase growth pattern. During the early phase, coverage expands rapidly as the fuzzer exercises shallow, easily-reachable code paths through its declaration and execution mutators. This phase saturates within the first few hundred iterations. In the later phase, growth slows and becomes sparse, punctuated by small step-increases as the scheduler promotes high-valuation seeds that unlock previously unreachable regions. This step behavior is clearly visible in Fig. 2, corresponding to a batch of mutator-generated scripts that first reach the override-dispatch paths in the object model.

The three runtimes differ substantially in their final coverage levels, reported here as the aggregate across multiple discrete run trials totaling up to 24 hours per runtime in Table IV, and per-file breakdowns are provided in Table VII, Table VIII, and Table IX.

**Table IV**
**Coverage Result**

| Targets | Functions Coverage (cov, %) | Lines Coverage (cov, %) | Branches Coverage (cov, %) |
|---|---|---|---|
| CPython | 4300, 37.45% | 66130, 25.69% | 23326, 19.00% |
| Lua | 762, 67.85% | 8595, 55.90% | 3390, 45.80% |
| QuickJS | 829, 48.71% | 15517, 31.75% | 6049, 25.81% |

Lua achieves the highest coverage, which reflects its comparatively compact codebase and the pervasive nature of its metamethod dispatch mechanism, and allows OverrideFuzz's override mutators to activate a large proportion of the runtime's C code paths. CPython's lower absolute percentages are expected, as CPython's instrumented codebase encompasses many modules (I/O, networking, ctypes) that are structurally unreachable from pure scripted inputs. QuickJS's coverage curve enters a plateau earlier than the other two runtimes, reflecting the dense concentration of executable paths in QuickJS's compact monolithic evaluation core.

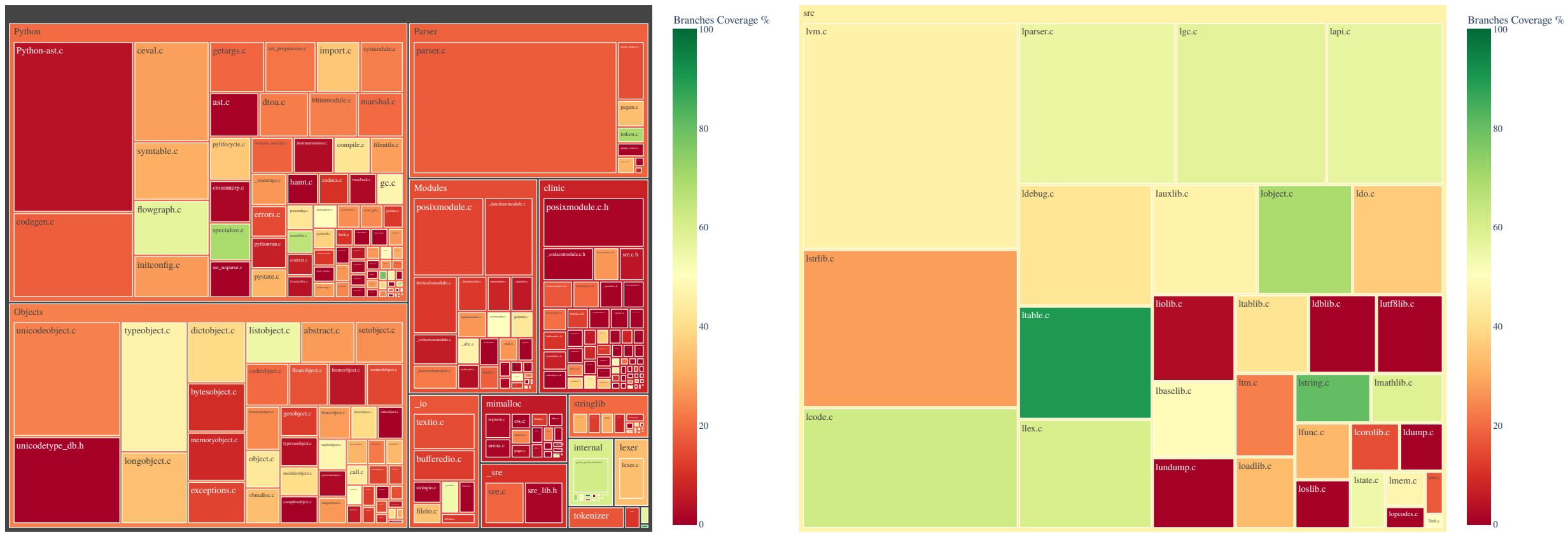

**Fig. 5. CPython per-file branches coverage treemap**

**Fig. 6. Lua per-file branches coverage treemap**

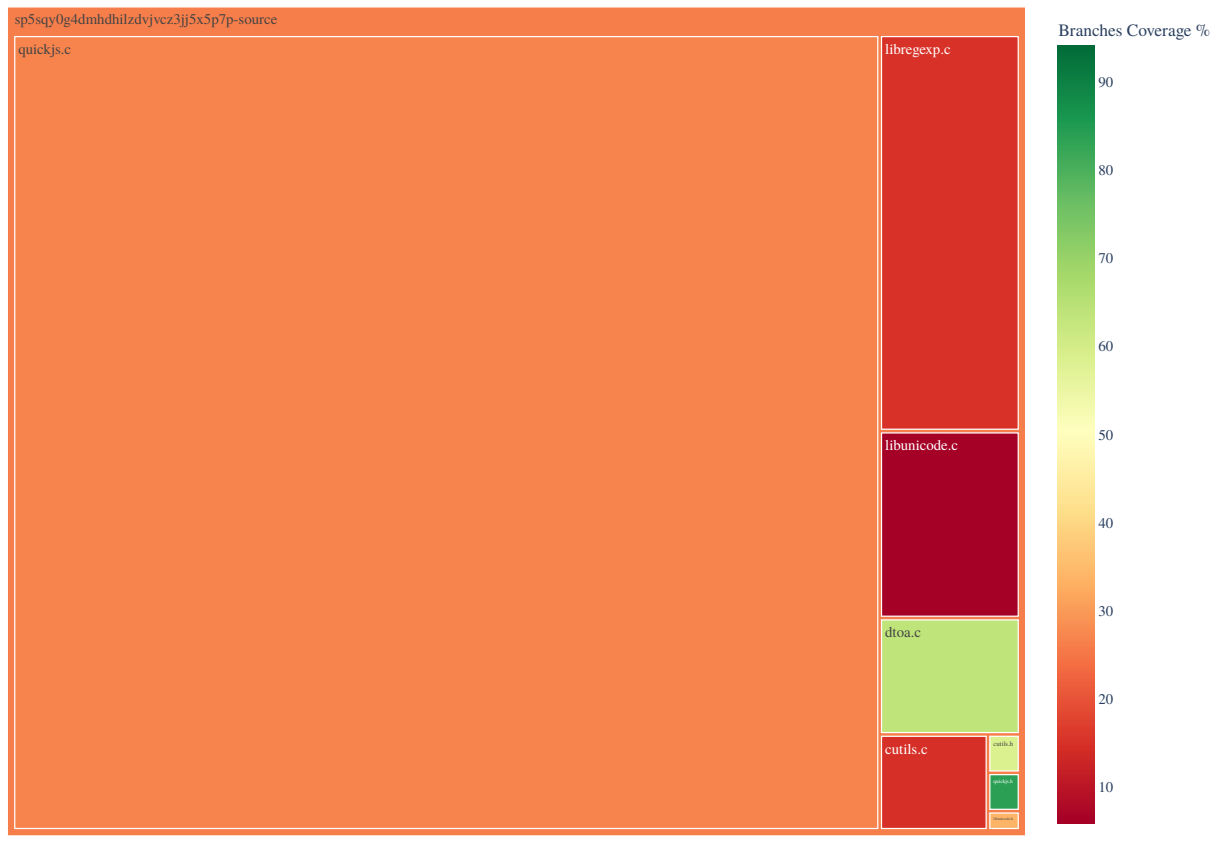

**Fig. 7. QuickJS per-file branches coverage treemap**

The treemap figures present the per-file coverage distribution, where each rectangle's area is proportional to its total instrumented code size and shading is proportional to the coverage achieved. For CPython, coverage concentrates in the object model and bytecode evaluation layers, while peripheral library modules remain sparsely covered. Lua's treemap shows broad and relatively uniform coverage across the metamethod dispatch and value-representation files, consistent with OverrideFuzz's design emphasis on override hook activation. QuickJS's treemap reflects its monolithic architecture, with the core evaluation file accounting for the majority of both code volume and covered branches.

Two design choices systematically depress the absolute coverage figures across all three runtimes. First, only a small subset of each runtime's built-in modules was enabled during evaluation to reduce noise from incidental module interactions, which proportionally reduces function coverage relative to the full instrumented codebase. Second, OverrideFuzz deliberately omits control-flow constructs such as loops and conditional branches from its execution mutator (see Section 3.3.2), because these constructs tend to stall execution or produce unreachable code without contributing meaningfully to override-dispatch path exploration. This omission depresses branch coverage independently of how broadly the fuzzer reaches into the codebase. Both choices are intentional and reflect the design priority of maximizing dispatch-path density over raw coverage breadth.

A direct quantitative comparison with prior work was not conducted for two reasons. First, works such as Reflecta [8] do not publish raw coverage measurements, and re-running those systems under controlled conditions was outside the time budget of this project. Second, OverrideFuzz evaluates CPython 3.14.3, whereas prior work evaluates different CPython versions, whose source layouts and instrumented line counts differ enough that coverage percentages are not directly comparable. Based on the CPython coverage treemap released with the Reflecta paper artifact, the prior result appears visually comparable to the treemap reported in this thesis. This observation supports only an informal comparison, and should be treated as indicative rather than conclusive.

## 4.4 Discovered Vulnerabilities

To answer RQ2, OverrideFuzz did not discover novel vulnerabilities in CPython, Lua, or QuickJS during the evaluation period (up to 24 hours per target across discrete trials). This outcome is primarily constrained by two practical factors: limited fuzzing time and engineering effort. In particular, the early termination problem described in Section 6.1 caused the fuzzer to discard entire generated function bodies whenever an early statement raised an exception, reducing the effective coverage signal per campaign hour. As a bachelor thesis without laboratory infrastructure or dedicated compute resources, the evaluation was conducted on a single workstation for a bounded campaign, which is insufficient to saturate the search space of override-dispatch interactions across three runtimes.

Despite the absence of new bug discoveries, examination of the saved corpus shows that OverrideFuzz does generate structurally relevant inputs. Corpus entries include subclasses of `bytes` class with overridden dunder methods, combined with bytearray-manipulating execution mutators (see Snippet I). The fuzzer independently reconstructs the patterns of inputs that triggered the motivation examples, but the specific interleaving of override invocation and internal allocator state required to exercise a window in which a freed pointer is held was not sampled within the available budget.

# CHAPTER 5

# CONCLUSION

Grammar and reflection-based fuzzers have substantially advanced script-language runtime testing, but they address syntactic validity and interface reachability rather than the semantic conditions under which a distinct class of vulnerabilities arises. That class consists of bugs triggered when user-defined hooks override built-in interpreter behavior with unexpected side effects. OverrideFuzz addresses this gap by combining a two-phase grammar model with active and passive reflection. The declaration phase constructs class hierarchies whose subclasses carry overriding methods, and the execution phase generates operations that route through those hooks, creating the interleaving of override invocation and interpreter-internal state transitions on which type-confusion and use-after-free bugs depend. The `globalRef` mechanism bridges the two phases by allowing override bodies to directly reference and mutate objects declared in enclosing scopes, enabling the fuzzer to construct precisely the kind of interleaving between override invocation and interpreter-internal allocator state that vulnerabilities such as those in Example I and Example II require.

OverrideFuzz achieves measurable and consistent coverage growth across CPython, Lua, and QuickJS, reaching 37.45%, 67.85%, and 48.71% function coverage respectively, with Lua benefiting most from the pervasive nature of its metamethod dispatch mechanism. No novel vulnerabilities were discovered within the bounded evaluation campaign, a result constrained by compute budget and engineering effort rather than by a fundamental design limit. Corpus analysis confirms that the fuzzer independently reconstructs structurally relevant inputs matching the pattern of the motivation examples.

OverrideFuzz further demonstrates that a language-portable, semantic-aware grammar fuzzer can be built from scratch without relying on an established fuzzing backend, and that adding a new target requires implementing only four driver components while others remain unchanged.

# CHAPTER 6

# FUTURE WORKS

## 6.1 Early Termination: Granularity of expressions in function

The current execution model generates a function body as a single, monolithic unit and submits the entire scope to the interpreter at once. Because OverrideFuzz cannot guarantee the semantic correctness of every statement within a generated body, a type mismatch or an undefined reference anywhere in the body causes the interpreter to raise an exception before the later statements execute. The entire generation effort yields no meaningful coverage signal, which is the early termination problem.

A finer-grained execution model addresses this directly. Rather than committing an entire raw scope to the interpreter, OverrideFuzz could wrap each line independently with a `try-except` or `if` block to prevent exceptions from aborting the entire body, or it could check each premise by inspecting the runtime state. Under this model, a failing statement affects only itself, similar to the outside line-by-line execution mode.

## 6.2 Architecture Limitations

The current architecture requires the target script-language interpreter to provide an embeddable API library. Some languages, which do not expose such an interface, cannot be supported by this platform directly. Furthermore, because OverrideFuzz is built entirely from scratch rather than layered atop an established fuzzing backend such as LibAFL or OSS-Fuzz, it does not inherit mature infrastructure for corpus distillation, coverage-guided feedback tuning, or parallel scaling.

## 6.3 Alternative: LLM-Assisted Auditing

The foundational premise of fuzzing is an economic one (as we discussed in Section 1): human auditing hours are scarce and expensive, so automated tools substitute cheap machine cycles for expert time. Recent advances in large language model (LLM) reasoning challenge this premise at the root. Rather than replacing man-hours with machine-hours, an LLM auditor can compress man-hours directly by reasoning semantically about source code, the submission logs, and inferring vulnerable patterns without wasting hours on trying to reach paths. Specifically, Anthropic's 2026 experiment applying Claude Opus 4.6 to open-source C projects demonstrated that an LLM can independently identify vulnerabilities [21], a capability that sidesteps the execution-volume assumption entirely. At the fuzzing systems level, DARPA's AI Cyber Challenge (AIxCC), which ran from 2023 to 2025, represents a large-scale trial of LLM systems design for cybersecurity that combine vulnerability discovery, fuzzing-style exploration, and automated remediation on real-world open-source software [22].

For a tool such as OverrideFuzz, this development carries a specific implication. The comparative advantage of semantic-aware grammar fuzzing is its ability to construct inputs that are both syntactically valid and semantically meaningful, inputs that a purely random fuzzer cannot generate. However, an LLM can read the interpreter source directly and reason toward the same edge cases, such as type confusion in override dispatch or reference-count errors in built-in methods, without requiring a mutation engine or a coverage signal. As LLM code reasoning over large C codebases matures, the machine-hour justification for grammar fuzzing weakens accordingly.

The practical near-term outcome is therefore a hybrid pipeline. Fuzzing and AI auditing are complementary rather than competing: a fuzzer can narrow the attack surface by identifying code regions with anomalous coverage growth, and an LLM auditor can then inspect those regions with semantic depth that exhaustive test generation cannot achieve. Taking this one step further, fuzzing can be applied within the auditing loop itself: wherever an LLM would otherwise iterate through candidate inputs at token cost, a fuzzer can substitute cheap machine cycles, reserving LLM inference for the semantic judgments that only language understanding can make. Under this hybrid model, traditional fuzzing remains useful, although its role shifts from the role of primary engine to behind the scene.

# APPENDIX

## A Override Functions Reference

The table below shows representative examples of override functions discovered by OverrideFuzz's active reflection phase for each supported runtime. The actual set is populated dynamically at startup by inspecting each interpreter's introspection API, and therefore varies across interpreter versions and build configurations.

**Table V**

**Representative override functions discovered by active reflection per target runtime (not exhaustive)**

| Runtime | Mechanism | Example Functions |
|---|---|---|
| CPython | Dunder methods on user-defined classes | `__index__`, `__getitem__`, `__setitem__`, `__getattr__`, `__setattr__`, `__eq__`, `__lt__`, `__add__`, `__new__`, `__init__`, `__del__`, `__len__`, `__hash__` |
| Lua | Metamethods via `setmetatable` | `__index`, `__newindex`, `__add`, `__sub`, `__eq`, `__lt`, `__call`, `__gc`, `__len`, `__tostring`, `__concat` |
| QuickJS | Prototype chain, property descriptors, and Proxy traps | `get`, `set`, `has`, `deleteProperty`, `apply`, `construct`, `Object.defineProperty`, `Symbol.toPrimitive`, `Symbol.iterator` |

## B Passive Reflection Correction Types

The table below lists representative error classes recognized by the passive reflection layer, the correction applied to the shared context model, and a representative CPython error message for each. The list is not exhaustive.

**Table VI**

**Representative passive reflection correction types with example CPython error messages (not exhaustive)**

| Error class | Correction applied | Example message (CPython) |
|---|---|---|
| Unary operator type mismatch | Remove offending type from the allowed type set for that operator | `bad operand type for unary ~: 'str'` |
| Missing attribute | Evict the attribute name from the property list of the identified type | `'dict' object has no attribute 'find'` |
| Wrong argument type | Update the parameter type at the faulting position in the stored function signature | `replace() argument 1 must be str, not None` |

| Error class | Correction applied | Example message (CPython) |
|---|---|---|
| Wrong argument count | Correct the stored arity of the function to match the interpreter's requirement | `iter() takes from 1 to 2 positional arguments but 0 were given` |

# C Per-File Coverage Breakdown

The tables below report LLVM source-based coverage aggregated from all run trials (up to 24 hours per target), extracted from the final merged `default.profdata` via `llvm-cov export`. Coverage is measured over the instrumented C source of each runtime. For CPython (Table VII), only source files with at least 400 covered lines are shown. Lua (Table VIII) and QuickJS (Table IX) show all instrumented source files with at least one percent covered line. Header-only files and files with no executable content are omitted.

**Table VII**

**CPython 3.14.3 per-file coverage (files with >=400 covered lines, sorted by function covered)**

| File | Functions (cov/tot, %) | Lines (cov/tot, %) | Branches (cov/tot, %) |
|---|---|---|---|
| errnomodule.c | 3/ 3,100.00% | 402/ 418, 96.17% | 144/ 288, 50.00% |
| assemble.c | 29/ 30, 96.67% | 465/ 568, 81.87% | 174/ 266, 65.41% |
| specialize.c | 50/ 55, 90.91% | 1155/ 1581, 73.06% | 462/ 668, 69.16% |
| flowgraph.c | 99/114, 86.84% | 1830/ 2912, 62.84% | 999/ 1762, 56.70% |
| compile.c | 47/ 59, 79.66% | 643/ 1240, 51.85% | 232/ 558, 41.58% |
| listobject.c | 98/124, 79.03% | 1596/ 2452, 65.09% | 545/ 994, 54.83% |
| tupleobject.c | 30/ 41, 73.17% | 413/ 717, 57.60% | 177/ 376, 47.07% |
| moduleobject.c | 32/ 44, 72.73% | 438/ 1029, 42.57% | 190/ 478, 39.75% |
| gc.c | 69/ 95, 72.63% | 797/ 1300, 61.31% | 168/ 358, 46.93% |
| dtoa.c | 20/ 28, 71.43% | 615/ 1690, 36.39% | 210/ 936, 22.44% |
| dictobject.c | 164/240, 68.33% | 2125/ 4211, 50.46% | 603/ 1546, 39.00% |
| longobject.c | 115/169, 68.05% | 1764/ 4053, 43.52% | 696/ 2072, 33.59% |
| import.c | 102/152, 67.11% | 1167/ 2658, 43.91% | 333/ 950, 35.05% |
| typeobject.c | 264/413, 63.92% | 3720/ 7736, 48.09% | 1720/ 3746, 45.92% |
| _warnings.c | 35/ 55, 63.64% | 414/ 1132, 36.57% | 129/ 488, 26.43% |
| symtable.c | 43/ 68, 63.24% | 891/ 2287, 38.96% | 572/ 1878, 30.46% |
| pylifecycle.c | 52/ 83, 62.65% | 901/ 1854, 48.60% | 269/ 764, 35.21% |
| unicodeobject.c | 225/371, 60.65% | 3300/10509, 31.40% | 1236/ 5380, 22.97% |
| descrobject.c | 55/ 94, 58.51% | 531/ 1084, 48.99% | 149/ 378, 39.42% |
| fileutils.c | 35/ 60, 58.33% | 477/ 1239, 38.50% | 141/ 508, 27.76% |
| bltinmodule.c | 39/ 69, 56.52% | 602/ 2036, 29.57% | 207/ 918, 22.55% |
| pystate.c | 75/137, 54.74% | 692/ 1447, 47.82% | 140/ 442, 31.67% |

| File | Functions (cov/tot, %) | Lines (cov/tot, %) | Branches (cov/tot, %) |
|---|---|---|---|
| abstract.c | 74/138, 53.62% | 640/ 2049, 31.23% | 266/ 984, 27.03% |
| setobject.c | 53/ 99, 53.54% | 651/ 1678, 38.80% | 223/ 876, 25.46% |
| initconfig.c | 72/137, 52.55% | 1120/ 2943, 38.06% | 366/ 1358, 26.95% |
| object.c | 54/109, 49.54% | 671/ 1464, 45.83% | 240/ 560, 42.86% |
| bytearrayobject.c | 51/104, 49.04% | 577/ 1685, 34.24% | 161/ 658, 24.47% |
| codegen.c | 76/161, 47.20% | 1280/ 4586, 27.91% | 765/ 4366, 17.52% |
| obmalloc.c | 56/122, 45.90% | 663/ 1726, 38.41% | 177/ 532, 33.27% |
| bytesobject.c | 43/ 94, 45.74% | 405/ 2301, 17.60% | 113/ 1242, 9.10% |
| ceval.c | 46/102, 45.10% | 613/ 2242, 27.34% | 892/ 3210, 27.79% |
| parser.c | 195/456, 42.76% | 8395/34902, 24.05% | 2157/12572, 17.16% |
| getargs.c | 21/ 52, 40.38% | 438/ 2186, 20.04% | 226/ 1202, 18.80% |
| codeobject.c | 29/ 94, 30.85% | 424/ 1998, 21.22% | 156/ 798, 19.55% |
| sysmodule.c | 36/131, 27.48% | 484/ 2055, 23.55% | 214/ 944, 22.67% |
| sre.c | 25/ 96, 26.04% | 404/ 1880, 21.49% | 204/ 1098, 18.58% |
| exceptions.c | 42/166, 25.30% | 417/ 2181, 19.12% | 142/ 1074, 13.22% |
| Python-ast.c | 34/166, 20.48% | 686/17052, 4.02% | 321/ 8748, 3.67% |
| _datetimemodule.c | 49/244, 20.08% | 579/ 4320, 13.40% | 194/ 1772, 10.95% |
| posixmodule.c | 38/288, 13.19% | 1014/ 5897, 17.20% | 345/ 2562, 13.47% |

**Table VIII**

**Lua 5.5.0 per-file coverage (files with >=1% covered lines, sorted by function covered)**

| File | Functions (cov/tot, %) | Lines (cov/tot, %) | Branches (cov/tot, %) |
|---|---|---|---|
| linit.c | 1/ 1,100.00% | 13/ 17, 76.47% | 3/ 6, 50.00% |
| ltable.c | 57/ 59, 96.61% | 700/ 766, 91.38% | 310/ 346, 89.60% |
| lobject.c | 24/ 25, 96.00% | 319/ 433, 73.67% | 165/ 240, 68.75% |
| lstring.c | 18/ 19, 94.74% | 203/ 225, 90.22% | 69/ 86, 80.23% |
| lmathlib.c | 31/ 33, 93.94% | 240/ 294, 81.63% | 48/ 82, 58.54% |
| lstate.c | 19/ 22, 86.36% | 236/ 275, 85.82% | 25/ 46, 54.35% |
| lvm.c | 27/ 32, 84.38% | 699/1309, 53.40% | 508/1118, 45.44% |
| lcode.c | 90/108, 83.33% | 832/1254, 66.35% | 369/ 596, 61.91% |
| lfunc.c | 14/ 17, 82.35% | 136/ 203, 67.00% | 23/ 72, 31.94% |
| ltablib.c | 14/ 17, 82.35% | 175/ 257, 68.09% | 55/ 132, 41.67% |
| lapi.c | 77/ 96, 80.21% | 790/1100, 71.82% | 242/ 432, 56.02% |
| lbaselib.c | 26/ 33, 78.79% | 241/ 379, 63.59% | 70/ 144, 48.61% |
| lcorolib.c | 11/ 14, 78.57% | 64/ 147, 43.54% | 8/ 54, 14.81% |
| lauxlib.c | 53/ 69, 76.81% | 478/ 744, 64.25% | 121/ 264, 45.83% |
| lmem.c | 6/ 8, 75.00% | 57/ 82, 69.51% | 15/ 32, 46.88% |
| lgc.c | 55/ 74, 74.32% | 641/1071, 59.85% | 318/ 556, 57.19% |

| File | Functions (cov/tot, %) | Lines (cov/tot, %) | Branches (cov/tot, %) |
|---|---|---|---|
| llex.c | 18/ 25, 72.00% | 293/ 435, 67.36% | 200/ 334, 59.88% |
| lstrlib.c | 52/ 73, 71.23% | 562/1254, 44.82% | 215/ 780, 27.56% |
| lparser.c | 75/107, 70.09% | 923/1528, 60.41% | 323/ 582, 55.50% |
| ldo.c | 30/ 43, 69.77% | 323/ 644, 50.16% | 82/ 228, 35.96% |
| ldebug.c | 29/ 49, 59.18% | 323/ 683, 47.29% | 156/ 372, 41.94% |
| loadlib.c | 14/ 27, 51.85% | 172/ 336, 51.19% | 32/ 98, 32.65% |
| ltm.c | 9/ 19, 47.37% | 93/ 235, 39.57% | 26/ 114, 22.81% |
| lzio.c | 2/ 5, 40.00% | 19/ 55, 34.55% | 3/ 18, 16.67% |
| liolib.c | 7/ 47, 14.89% | 49/ 450, 10.89% | 4/ 164, 2.44% |
| lutf8lib.c | 1/ 12, 8.33% | 6/ 190, 3.16% | 0/ 120, 0.00% |
| loslib.c | 1/ 19, 5.26% | 4/ 197, 2.03% | 0/ 62, 0.00% |
| ldblib.c | 1/ 28, 3.57% | 4/ 323, 1.24% | 0/ 122, 0.00% |

**Table IX**

**QuickJS 2025-09-13 per-file coverage (files with >=1% covered lines, sorted by function covered)**

| File | Functions (cov/tot, %) | Lines (cov/tot, %) | Branches (cov/tot, %) |
|---|---|---|---|
| dtoa.c | 38/ 40, 95.00% | 873/ 1045, 83.54% | 303/ 478, 63.39% |
| quickjs.c | 675/1434, 47.07% | 13530/42898, 31.54% | 5357/20170, 26.56% |
| libregexp.c | 27/ 59, 45.76% | 548/ 2422, 22.63% | 243/ 1628, 14.93% |
| cutils.c | 11/ 31, 35.48% | 116/ 457, 25.38% | 44/ 304, 14.47% |
| libunicode.c | 8/ 47, 17.02% | 119/ 1597, 7.45% | 45/ 768, 5.86% |

# D Representative Corpus Samples

The following samples are taken directly from the saved corpus produced by OverrideFuzz on CPython. Method bodies are condensed to highlight structural features. All samples illustrate the two-phase structure of generated programs: a declaration phase that initializes shared mutable state and defines override-bearing classes, followed by an execution phase inside each override method that applies mutations to the shared context.

```
# scope 0
str_a = "?H0o\\\\\\x\\w\\\\\\\\\\\\\\\\\\\\\\\\\\\\\\\\\\\\\\\\\\\\\\\\\\\\\\\\\\\\\\\\\\\\\\\\
\\\\\\\\\\\\\\1\\\\\\\\" # havoc
str_b = "u\\\\D\\z\\\\\\\\\\\\\\\\\\\\\\\\\\\\\\\\\\\\\\\\\\\\\\\\\\\\\\\\\\\\\\\\\\\\\\\\\\\\\\\\\\
\s\\\\\\\\\\\\\\\\\\\\\\\\"
byte_a = b""
int_a = 831522344197812962
int_b = -17
float_a = 414132
bool_a = False
bool_b = False
list_a = [3, 1, 4, 1, 5]
ba_a = bytearray(56)
dict_a = {}
class aaa(bytes):                  # bytes subclass
    def rstrip(arg_a, arg_b):      # overrides bytes.rstrip
        # scope 1
        exec('from math import *', globals())
        z50 = bytes()
        # ... declaration mutation continue ...
        global second, _use_args_, dict_a... # globalRef
        # Execution mutations: bytearray indexed writes inside override body
        l82.__path__[l10] = z44
        ah51 = tzinfo.fromutc()
        ah52 = minute()
        aao[day.nan] = e58
        # ... execution mutations continue ...
    def removeprefix(arg_a, arg_b):
        # scope 2
        exec('from json import *', globals())
        # ... declaration continue ...
        global CODESIZE, z46, utc, denominator ... # globalRef
        ah79 = resolution.__mul__()
        l08[min.__radd__(s33)] = ad7
        max.month = byte_a.rstrip(date)
        ah80 = __version__.find(days.fma(max, microseconds, __traceback__),
aa72.microsecond, min.second)
        ah81 = JSONEncoder(s33, s33, s33, e14, s33, min.hour, max.resolution,
max.resolution)
        # ... execution mutations continue ...
# ... declaration and execution mutation continue ...
```

**Snippet I. Corpus sample: bytes subclass with bytearray subscript writes inside an override method (from corpus/saved/2.py)**

# E Sanitizer Coverage instrumentation Code

```
extern uint32_t newEdgeCnt;
extern "C" void __sanitizer_cov_trace_pc_guard_init(uint32_t *start,
                                                    uint32_t *stop) {
    if (start == stop || *start)
        return;
    static uint32_t N = 0;
    for (uint32_t *x = start; x < stop; ++x) {
        *x = ++N;
    }
}

extern "C" void __sanitizer_cov_trace_pc_guard(uint32_t *guard) {
    if (!*guard)
        return;
    newEdgeCnt++;
    *guard = 0;
}
```

**Snippet II. OverrideFuzz sanitizer coverage instrumentation code**